\documentclass[aps,prfluids,onecolumn,showpacs,showkeys,
longbibliography,superscriptaddress,amsmath,amssymb,floatfix,nofootinbib,a4paper]{revtex4-2}

\usepackage{mathtools}
\usepackage{graphicx}
\usepackage[usenames,dvipsnames]{color}
\usepackage{xcolor}
\usepackage{epsfig}
\usepackage{wasysym}
\usepackage{natbib}
\usepackage{bm}
\usepackage[normalem]{ulem}

\newcommand{\br}{{\bf r}}

\newcommand{\be}{{\bf e}}
\newcommand{\bn}{{\bf n}}
\newcommand{\bk}{{\bf k}}

\newcommand{\bs}{{\bf s}}

\newcommand{\bu}{{\bf u}}
\newcommand{\bV}{{\bf V}}
\newcommand{\bA}{{\bf A}}
\newcommand{\bB}{{\bf B}}

\newcommand{\bQ}{{\bf Q}}
\newcommand{\bF}{{\bf F}}

\newcommand{\bK}{{\bf K}}

\newcommand{\bT}{{\bf T}}

\newcommand{\bOmega}{{\boldsymbol{\Omega}}}

\newcommand{\bgamma}{{\boldsymbol{\gamma}}}

\newcommand{\Spart}{{\mathcal{S}_p}}
\newcommand{\brp}{{\br_p}}
\newcommand{\Qpot}{{\Phi}}
\newcommand{\intfluid}{\smashoperator{\int_{\mathcal{D}}} d\mathcal{V}\;}
\newcommand{\intinfty}{\smashoperator{\oint_{\mathcal{S}_\infty}} d\boldsymbol{\mathcal{S}}}
\newcommand{\intwall}{\smashoperator{\oint_\Spart} d\boldsymbol{\mathcal{S}}}

\newcommand{\eq}[1]{Eq.~(\ref{#1})}
\newcommand{\eqs}[1]{Eqs.~(\ref{#1})}
\newcommand{\fig}[1]{Fig.~\ref{#1}}
\newcommand{\rcite}[1]{Ref.~\cite{#1}}
\newcommand{\eref}[1]{(\ref{#1})}

\newcommand{\Fref}[1]{Figure~\ref{#1}}
\def\bp{\!\!\!}

\begin{document}

\title{Force--dependence of the rigid--body motion for an
  arbitrarily shaped 
  particle \\
  in a forced, incompressible Stokes flow
  }

\author{Alvaro Dom\'\i nguez}
\email{\texttt{dominguez@us.es}}
\affiliation{F\'\i sica Te\'orica, Universidad de Sevilla, Apdo.~1065, 
41080 Sevilla, Spain}
\affiliation{Instituto Carlos I de F{\'i}sica Te{\'o}rica y
  Computacional, 18071 Granada, Spain}

\author{Mihail N. Popescu}
\email{\texttt{mpopescu@us.es}}
\affiliation{F\'\i sica Te\'orica, Universidad de Sevilla, Apdo.~1065, 
41080 Sevilla, Spain}


\begin{abstract}
  When a particle moves in a Newtonian flow at low Reynolds number,
  inertia is irrelevant and a linear relationship exists between
  velocities and forces. For incompressible flows, any force
  distribution $\mathbf{f}(\br)$ acting in the fluid bulk induces flow
  and motion only through its solenoidal component.  For force
  distributions that are spatially localized (i.e., vanish
  sufficiently fast at infinity), we derive the representation of the
  rigid body motion as an explicit linear functional of
  $\nabla\times\mathbf{f}$, which complements the usual representation
  in terms of $\mathbf{f}$. We illustrate the utility of this
  alternative representation, which has the advantage of having the
  incompressibility constraint built-in, in avoiding certain
  ambiguities that arise, e.g., when implementing approximations for
  swimmers.
\end{abstract}

\maketitle

When dealing with the motion of rigid particles within a Newtonian
fluid flow at low Reynolds number, the Lorentz reciprocal theorem
\cite{Lorentz_original,Lorentz_transl,HaBr73,KiKa91,MaSt19} for the
unforced Stokes equations is a useful tool for extracting relevant
information about the motion of the particle while sidestepping the
explicit calculation of the flow. This theorem can be generalized
\cite{KiKa91,Teub82} to account for a bulk force field
$\mathbf{f}(\br)$ acting in the fluid: notable applications are the
derivation of Fax{\'e}n laws for particles of arbitrary shape, see,
e.g., \cite[][Ch. 3]{KiKa91}, and the calculation of the translational
and rotational velocities of a self-phoretic particle, see, e.g.,
Refs.~\cite{SaSe12,Brown17,DPRD20,DAPA20,ANV22,ShOl24,DoPo24}. 
The incompressibility constraint is enforced through the hydrodynamic
pressure, that also adsorbs any potential (longitudinal) component of
the bulk force field; therefore only the solenoidal (transversal)
component may induce flow and motion of the particle.  We revisit here
the generalized reciprocal theorem and formulate it in terms of
$\nabla\times\mathbf{f}(\br)$, which renders a motion--force
relationship with the incompressibility constraint explicitly
accounted for. We show, as an example, that this formulation is more advantageous to
use in the configuration, often occurring in phoresis and
self-phoresis, that the effect of the force is relevant only within a
thin layer region near the particle.

\begin{figure}[!t]
  \hfill
  \begin{tabular}[c]{c}
    \includegraphics[width=.4\columnwidth]{
    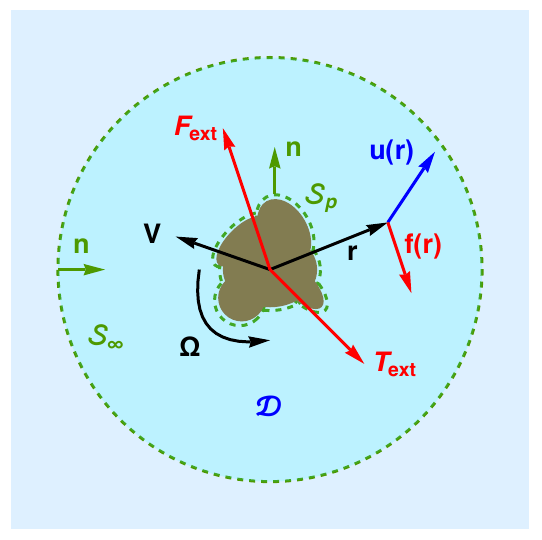
    }
  \end{tabular}
  \hfill
  \begin{tabular}[c]{c}
    \includegraphics[width=.4\columnwidth]{
    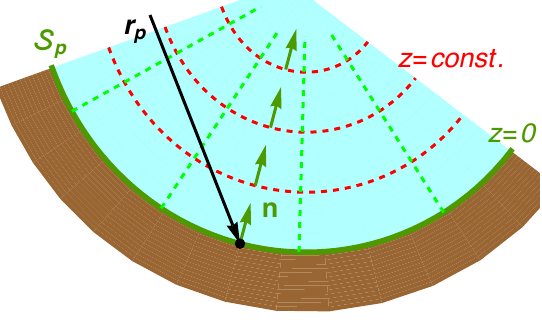
    }
  \end{tabular}
  \hspace*{\fill}
  \caption{\textit{(Left)} Schematic description
    of 
    the system under consideration: a rigid, impermeable particle
    moves with translational and rotational velocities $\bV$ and
    $\bOmega$, respectively, in a Stokes flow $\bu(\br)$ defined in
    the fluid domain $\mathcal{D}$ bounded by the surface of the
    particle, $\Spart$, and a surface at infinity,
    $\mathcal{S}_\infty$ (both oriented into the fluid domain, as
    indicated by the unit normal $\bn$). There is a force,
    $\mathbf{F}^\mathrm{(p)}_\mathrm{ext}$, and a torque,
    $\mathbf{T}^\mathrm{(p)}_\mathrm{ext}$, of external origin acting
    directly on the particle, and a force field $\mathbf{f}(\br)$
    acting in the bulk of the fluid.
    \textit{(Right)} Definition of a local system of
    coordinates 
    near the surface of the particle. Here, $\brp$ is an arbitrary
    point of the particle's surface $\Spart$, and $\bn(\brp)$ is the
    unit normal at that point.
    Near the surface of the particle,
    any point in the fluid
    can be parametrized as $\br=\brp+z\bn(\brp)$ with $z\geq 0$
    being the distance from the surface along the normal, and any
    vector $\bQ(\br = \brp + z \bn)$ can be decomposed locally into
    its normal and tangential components as
    $\bQ = \bn Q_z + \bQ_\parallel$.
    }
  \label{fig:schematic}
\end{figure}

Consider an arbitrarily shaped, rigid, and impermeable particle
immersed in a Newtonian fluid. The particle translates with velocity
$\bV$ and rotates with angular velocity $\bOmega$, while the fluid
flows with the velocity field $\bu(\br)$ in the domain $\mathcal{D}$,
see \Fref{fig:schematic}.  This flow (i) is incompressible, (ii) obeys
the Stokes equations with forcing in the bulk of the fluid, (iii)
satisfies a no--slip condition on the surface of the particle
$\Spart$, which we take oriented into the fluid, and (iv) vanishes at
infinity, which sets the rest frame with respect to which the particle
velocity is measured. In terms of the stress tensor
\begin{equation}
  \label{eq:stress}
  \mathsf{\Pi} = \eta \, [ \nabla\bu + (\nabla\bu)^\dagger ] -
  \mathsf{I} \, p ,
\end{equation}
that involves the viscosity $\eta$ and the hydrodynamic pressure $p$, this physical problem is phrased as
the following boundary--value problem for the fluid flow $\bu(\br)$:
\begin{subequations}
  \label{eq:bvp}
\begin{equation}
  \label{eq:stokes}
  \nabla\cdot\mathsf{\Pi}(\br) + \mathbf{f}(\br) = 0 ,
  \qquad
  \nabla\cdot\bu(\br)=0 ,
  \qquad
  \br\in\mathcal{D} ,
\end{equation}
\begin{equation}
  \label{eq:noslip}
  \bu(\br) =\bV + \bOmega\times\br, 
  \qquad
  \br\in\Spart ,
\end{equation}
\begin{equation}
  \label{eq:bcinfty}
  \bu(\br) \to 0 ,
  \qquad
  \mathrm{as}\; |\br|\to \infty .
\end{equation}
\end{subequations}
Here, the field $\mathbf{f}(\br)$ represents a given force density
acting on the fluid; this force is sourced either by external fields
or by interactions of the fluid constituents with the particle (e.g.,
an adsorption potential extending into the fluid), so that one writes
 \begin{equation}
  \label{eq:force}
  \mathbf{f}(\br) = \mathbf{f}_\mathrm{ext}(\br) + \mathbf{f}_\mathrm{part}(\br) .
\end{equation}
We assume that each of these two components has a compact support or, more
generally, that it vanishes at infinity as fast as needed. 
The particle immersed in the fluid is thus accordingly acted by the
reaction to the force field $\mathbf{f}_\mathrm{part}$ and by the
hydrodynamic stresses due to the motion relative to the
fluid;\footnote{The particle does not react to
      $\mathbf{f}_\mathrm{ext}$ by the very definition of the latter. 
      } 
in addition, there can be a force $\bF^{\mathrm{(p)}}_\mathrm{ext}$ and a torque
$\bT^{\mathrm{(p)}}_\mathrm{ext}$ due to external fields acting directly on the particle. (In general, the external fields acting on the particle and those responsible of
    $\mathbf{f}_\mathrm{ext}$ are different.)

In the limit of negligible inertia (overdamped particle motion and
Stokes flow), forces and torques are balanced, so that mechanical
balance (equilibrium) of the composed system ``particle~+~fluid'' 
is expressed in terms of the hydrodynamic stresses transmitted by the
fluid through a surface $\mathcal{S}_\infty$ at infinity (also taken to be oriented into the fluid):
\begin{equation}
\label{eq:balance_system}
  \bF^{\mathrm{(p)}}_\mathrm{ext}
  + 
  \bF^\mathrm{(f)}_\mathrm{ext}
  - \intinfty \cdot \mathsf{\Pi}
  = 0 ,
  \qquad
  \bT^\mathrm{(p)}_\mathrm{ext}
  + 
  \bT^\mathrm{(f)}_\mathrm{ext}
  - \oint_{\mathcal{S}_\infty} \bp\bp \br\times \left(
    d\boldsymbol{\mathcal{S}}\cdot \mathsf{\Pi} \right)
  = 0 \,,
\end{equation}
where
\begin{equation}
  \label{eq:FTextfluid}
  \bF^\mathrm{(f)}_\mathrm{ext} := \intfluid \mathbf{f}_\mathrm{ext} ,
  \qquad
  \bT^\mathrm{(f)}_\mathrm{ext} := \intfluid \br\times\mathbf{f}_\mathrm{ext}
\end{equation}
denote the total force and torque, respectively, acting on the fluid due to the force
density $\mathbf{f}_\mathrm{ext}$. 
When these expressions
are combined with the integrated version of the force balance on the
fluid derived from  \eq{eq:stokes}, one obtains the force
and torque balance on the particle:
\begin{equation}
  \label{eq:balance_particle}
  \bF^\mathrm{(p)}_\mathrm{ext}
  - \intfluid \mathbf{f}_\mathrm{part}
  + \intwall \cdot \mathsf{\Pi}
  = 0 ,
  \qquad
  \bT^\mathrm{(p)}_\mathrm{ext}
   - \intfluid \br\times\mathbf{f}_\mathrm{part} 
   + \smashoperator{\oint_{\Spart}} 
   \br\times \left(
    d\boldsymbol{\mathcal{S}}\cdot \mathsf{\Pi} \right)
  = 0 .
\end{equation}
These expressions will provide the (linear) relationship between the forces and torques
($\bF^\mathrm{(p,f)}_\mathrm{ext}, \bT^\mathrm{(p,f)}_\mathrm{ext},
\mathbf{f}$) and the velocities ($\bV, \bOmega$) (see, e.g.,
Refs. \cite{HaBr73,KiKa91,Teub82,DPRD20} and \eqs{eq:VOlin}
below). The pressure $p$ in \eq{eq:stress} plays the role of a
constraining force that enforces the incompressibility condition
$\nabla\cdot\bu=0$. Any additive gradient in the bulk force
$\mathbf{f}$ is thus irrelevant for the flow and for the motion of the particle, as
it can be absorbed in the definition of the pressure. That is, only
the solenoidal component of $\mathbf{f}$ is relevant, and therefore it
must be possible to find a result that involves just the curl of
this field. The goal in the following is
to find such expressions for $\bV$ and $\bOmega$, which complement the
framework discussed above and could present advantages in numerical
studies, as the irrelevance of the potential component of $\mathbf{f}$
will be explicitly  accounted for by
construction.

In the absence of any force density acting in the bulk of the fluid,
i.e., $\mathbf{f}_\mathrm{ext}(\br) \equiv 0$ and
$\mathbf{f}_\mathrm{part}(\br) \equiv 0$, one recovers the standard
problem of flow and particle motion driven by the direct action of an
external field on the particle \cite{HaBr73,KiKa91}. We denote the solution of this problem with
  primed symbols:
\begin{subequations}
  \label{eq:bvp_aux}
\begin{equation}
  \label{eq:stokes2}
  \nabla\cdot\mathsf{\Pi}'(\br) = 0 ,
  \qquad
  \nabla\cdot\bu'(\br)=0 ,
  \qquad
  \br\in\mathcal{D} ,
\end{equation}
\begin{equation}
  \label{eq:noslip2}
  \bu'(\br) =\bV' + \bOmega'\times\br, 
  \qquad
  \br\in\Spart ,
\end{equation}
\begin{equation}
  \label{eq:bcinfty2}
  \bu'(\br)  \to 0 ,
  \qquad
  \mathrm{as}\; |\br|\to \infty .
\end{equation}
\end{subequations}
with the additional relations (\eqs{eq:balance_system} and
\eref{eq:balance_particle} with $\mathbf{f}_\mathrm{ext}(\br)\equiv 0$
and $\mathbf{f}_\mathrm{part}(\br)\equiv 0$)
\begin{subequations}
\label{eq:mechIsol2}
\begin{equation}
  \label{eq:mechIsolationF2}
  {\bF'}^\mathrm{(p)}_\mathrm{ext} - \intinfty \cdot \mathsf{\Pi}' = 0 ,
  ~~
  {\bT'}^\mathrm{(p)}_\mathrm{ext}
  - \smashoperator{\oint_{\mathcal{S}_\infty}} 
  \br\times \left(
    d\boldsymbol{\mathcal{S}}\cdot \mathsf{\Pi}' \right)
  = 0 ,
\end{equation}
\begin{equation}
  \label{eq:hydroF2}
  {\bF'}^\mathrm{(p)}_\mathrm{ext} + \intwall \cdot \mathsf{\Pi}' = 0, 
  ~~
  {\bT'}^\mathrm{(p)}_\mathrm{ext}
  +  \smashoperator{\oint_{\Spart}} 
  \br\times \left(
    d\boldsymbol{\mathcal{S}}\cdot \mathsf{\Pi}' \right)
  = 0 .
\end{equation}
\end{subequations}
In this case the velocities of rigid body motion can be expressed as
the following linear combination of the external forces and torques
acting on the particle \cite{HaBr73,KiKa91}:
\begin{subequations}
\label{eq:rigbod_mobility}
\begin{equation}
  \label{eq:mobilityV}
  \bV' = \frac{1}{\eta} \mathsf{M}_t\cdot\bF'_\mathrm{ext}
  + \frac{1}{\eta} \mathsf{M}_{c} \cdot \bT'_\mathrm{ext} ,
\end{equation}
\begin{equation}
  \label{eq:mobilityO}
  \bOmega' = \frac{1}{\eta} \mathsf{M}_{c}^+ \cdot\bF'_\mathrm{ext}
  + \frac{1}{\eta} \mathsf{M}_{r} \cdot \bT'_\mathrm{ext} ,
\end{equation}
\end{subequations}
in terms of the 2nd-rank mobility tensors for translation,
$\mathsf{M}_t$, rotation, $\mathsf{M}_r$, and 
cross--coupling,
$\mathsf{M}_c$. These tensors are determined solely by the shape of
the particle; the first two are symmetric, the third one is not, in
general:
\begin{equation}
  \label{eq:symM}
  \mathsf{M}_t^+ = \mathsf{M}_t ,
  \qquad
  \mathsf{M}_r^+ = \mathsf{M}_r .
\end{equation}
Furthermore, in view of the linearity of the boundary--value problem,
the flow field in this case can be written formally as 
\begin{equation}
  \label{eq:uprime}
  \bu'(\br) = \bV' \cdot \left[ \mathsf{I} + \mathsf{K}(\br) \right]
  + \bOmega'\times\left[ \br + \bK(\br) \right] ,
  \quad
  \br\in\mathcal{D} ,
\end{equation}
where $\mathsf{I}$ is the identity 2nd-rank tensor, while the vector
field $\bK(\br)$ and the 2nd-rank tensor field $\mathsf{K}(\br)$ also
depend --- similarly to the mobility tensors --- only on the shape of
the particle \cite{HaBr73}. (Notice that the tensor $\mathsf{K}$ is
not symmetric in general, so that the order in which contractions with
it are written is important.\footnote{We note on passing that the
  difference between covariant and contravariant vectors is neglected
  in view that one can work in a global Cartesian coordinate system.})
From the boundary conditions, Eqs.~(\ref{eq:noslip2},
\ref{eq:bcinfty2}), obeyed by $\bu'(\br)$ it follows that the vector
fields $\bK(\br)$ and $\mathsf{K}(\br)$ satisfy
\begin{subequations}
\label{eq:BCsG}
\begin{equation}
  \label{eq:GatPart}
  \mathsf{K}(\br) = 0 ,
  \;\;
  \bK(\br) = 0 ,
  \qquad
  \br\in\Spart ,
\end{equation}
\begin{eqnarray}
  \label{eq:GatInfty}
  \mathsf{K}(\br)
  & \to
  & -\mathsf{I}
    + \textrm{multipolar    expansion},\qquad
  \mathrm{as}\; |\br|\to \infty \,\nonumber\\
  \;\;
  \bK(\br)
  & \to
  & - \br
    + \textrm{multipolar    expansion},
  \qquad
  \mathrm{as}\; |\br|\to \infty \,,
\end{eqnarray}
\end{subequations}
where the multipolar expansion accounts for all the fundamental
  singularities that describe the induced flow. Meanwhile, the
incompressibility constraint $\nabla\cdot\bu'=0$ leads straightforwardly to
\begin{equation}
  \label{eq:Gincomp}
  \nabla\cdot \left[ \be_j\cdot\mathsf{K}(\br) \right] = 0 ,
  \qquad 
  \nabla\times \bK(\br) = 0,
\end{equation}
where the unit constant vectors $\be_j$ (with $j=1,2,3$) form a Cartesian coordinate system. Consequently, one can write the fields $\bK(\br)$ and
$\mathsf{K}(\br)$ in terms of scalar and vector potentials,
respectively, as\footnote{\label{foot:A} One could introduce a
  2nd-rank tensor field $\mathsf{A} := \sum_j \bA^{(j)} \be_j$ and
  write $\mathsf{K}^\dagger = \nabla\times\mathsf{A}$ more compactly,
  but we prefer to work with the three vector fields $\bA^{(j)}$ for
  reasons of clarity.}
\begin{subequations}
  \label{eq:potentials}
\begin{equation}
    \label{eq:potA}
  \be_j\cdot\mathsf{K}(\br) = \nabla\times \bA^{(j)} (\br) ,
   \qquad
   j=1,2,3,
\end{equation}
\begin{equation}
  \label{eq:potQ}
  \bK(\br) = \mbox{} - \nabla \Qpot(\br) .
\end{equation}
\end{subequations}
Additionally, the potentials must satisfy boundary conditions,
following from \eqs{eq:BCsG}, at the surface of the particle and at
infinity; the details are provided in the Supplementary Material
\cite{SM}, and here we only note that, on the surface of the particle,
the scalar potential must be constant and the vector potentials must
be perfect gradients. Furthermore, one notes that \eqs{eq:potentials}
define the potentials up to a gauge transform,
\begin{equation}
  \label{eq:gauge}
  \Qpot(\br) \mapsto \Qpot(\br) + \phi,
  \qquad
  \bA^{(j)}(\br) \mapsto \bA^{(j)}(\br) + \nabla\alpha^{(j)}(\br) .
\end{equation}
The constant $\phi$ and the scalar field $\alpha(\br)$ are not
completely arbitrary because they must also comply with the boundary
conditions; it turns out that by providing these fields as boundary
conditions at the surface of the particle, i.e.,
\begin{equation}
  \label{eq:AQatPart}
  \Qpot(\br) = \phi ,
  \quad
  \bA^{(j)}(\br) = \nabla \alpha^{(j)}(\br) ,  
  \qquad
  \br\in\Spart , \quad j=1,2,3 ,
\end{equation}
and requiring that $\nabla^2 \alpha^{(j)}(\br) = \nabla\cdot\bA^{(j)}$
fixes the gauge \cite{SM}. A specific example, which will be
particularly useful in the following, is the ``Coulomb gauge'',
defined by the choice
\begin{equation}
  \label{eq:Coulomb}
  \phi=0,
  \qquad
  \alpha^{(j)}(\br\in\mathcal{D})\equiv 0 ,
  \quad j=1,2,3 .
\end{equation}
Accordingly, \eqs{eq:gauge} and \eref{eq:AQatPart} can be understood
as the rule for constructing the potentials in any gauge by starting
from the ones in the Coulomb gauge. In other words, the gauge freedom
is exhausted by providing the surface fields and the values of the
divergence for the vector potentials.

Returning to the general case $\mathbf{f} \neq 0$, the dependence of
the velocities on forces and torques can be derived (while
sidestepping the explicit calculation of the velocity field
$\bu(\br)$) by applying the Lorentz reciprocal theorem
\cite{Lorentz_original,Lorentz_transl} to the two hydrodynamic problems introduced in the
previous sections, namely the Stokes flows with (unprimed quantities) and without 
(primed quantities) the bulk force $\mathbf{f}(\br)$. The following relationship 
holds (the presence of the bulk force $\mathbf{f}$ in \eq{eq:stokes} does not change the
standard reasoning presented in, e.g., Ref.~\cite{HaBr73}; see also Refs.~\cite{KiKa91,Teub82} and \cite[Supp.~Mat.]{DPRD20}):
\begin{equation}
  \label{eq:reciprocal}
  \smashoperator{\oint_{\Spart\cup \mathcal{S}_\infty}}
  d\boldsymbol{\mathcal{S}}\cdot \mathsf{\Pi}\cdot\bu' 
  - \intfluid \mathbf{f}\cdot\bu'
  = \smashoperator{\oint_{\Spart\cup \mathcal{S}_\infty}}
  d\boldsymbol{\mathcal{S}}\cdot \mathsf{\Pi}'\cdot\bu ,
\end{equation}
after proper account of the orientation chosen for the surfaces
$\Spart$ and $\mathcal{S}_\infty$. Now, the integrals over
$\mathcal{S}_\infty$ vanish due to the boundary
conditions~(\ref{eq:bcinfty}, \ref{eq:bcinfty2}) and the fact that
  the force field $\mathbf{f}$ vanishes fast at infinity (so that
  the flows and the stress tensors
  decay at least as $1/r$ and $1/r^2$, respectively). The integrals
over $\Spart$ can be simplified by applying the boundary
conditions~(\ref{eq:noslip}, \ref{eq:noslip2}) and the force
balances~(\ref{eq:balance_particle}, \ref{eq:hydroF2}), so that \eq{eq:reciprocal} takes the form:
  \begin{equation}
    \bV' \cdot \left( - \bF^\mathrm{(p)}_\mathrm{ext}
      + \intfluid \mathbf{f}_\mathrm{part}
    \right)
    + \bOmega' \cdot \left( - \bT^\mathrm{(p)}_\mathrm{ext}
    + \intfluid \br\times\mathbf{f}_\mathrm{part} \right)
  - \intfluid \bu'\cdot \mathbf{f}
  = - {\bF'}^\mathrm{(p)}_\mathrm{ext} \cdot \bV
  - {\bT'}^\mathrm{(p)}_\mathrm{ext} \cdot \bOmega .
\end{equation}
By inserting  the representation~(\ref{eq:uprime}) for the velocity field
$\bu'(\br)$ in the equation above, one gets
\begin{equation}
  \bV' \cdot \left(
    \bF^\mathrm{(p)}_\mathrm{ext}
    + \intfluid \mathbf{f}_\mathrm{ext}
    + \intfluid \mathsf{K} \cdot \mathbf{f}
  \right)
  + \bOmega' \cdot \left( \bT^\mathrm{(p)}_\mathrm{ext}
    + \intfluid \br\times\mathbf{f}_\mathrm{ext}
    + \intfluid \bK\times\mathbf{f}
  \right)
  = {\bF'}^\mathrm{(p)}_\mathrm{ext} \cdot \bV
  + {\bT'}^\mathrm{(p)}_\mathrm{ext} \cdot \bOmega. 
\end{equation}
Finally, by using the definitions~(\ref{eq:FTextfluid}),
inserting the transposed version of
relations~(\ref{eq:rigbod_mobility}) simplified by \eq{eq:symM}, and noting that ${\bF'}^\mathrm{(p)}_\mathrm{ext}$, ${\bT'}^\mathrm{(p)}_\mathrm{ext}$ are independent and arbitrary, one arrives at the following result:
\begin{subequations}
  \label{eq:VOlin}
\begin{equation}
  \label{eq:V}
  \bV = \frac{1}{\eta} \mathsf{M}_t \cdot 
  \left( 
  \bF^\mathrm{(p)}_\mathrm{ext}
    + \bF^\mathrm{(f)}_\mathrm{ext}
    + \bF_\mathrm{mot}
  \right)
  + \frac{1}{\eta} \mathsf{M}_c \cdot 
  \left(
    \bT^\mathrm{(p)}_\mathrm{ext} 
    + \bT^\mathrm{(f)}_\mathrm{ext}
    + \bT_\mathrm{mot}
   \right) ,
\end{equation}
\begin{equation}
  \label{eq:O}
  \bOmega
  =
  \frac{1}{\eta}\mathsf{M}_c^+ \cdot 
  \left(
    \bF^\mathrm{(p)}_\mathrm{ext}
    + \bF^\mathrm{(f)}_\mathrm{ext}
    + \bF_\mathrm{mot}
  \right)
  + \frac{1}{\eta} \mathsf{M}_r \cdot 
  \left(
    \bT^\mathrm{(p)}_\mathrm{ext} 
    + \bT^\mathrm{(f)}_\mathrm{ext}
    + \bT_\mathrm{mot}
  \right) ,
\end{equation}
\end{subequations}
where we have defined
\begin{equation}
  \label{eq:FTswim}
  \bF_\mathrm{mot}
  := \intfluid
  \mathsf{K}(\br) \cdot \mathbf{f}(\br) ,
  \qquad 
  \bT_\mathrm{mot}
  := \intfluid
  \bK(\br) \times\mathbf{f}(\br) .
\end{equation}
These latter quantities encode the flow--mediated effect of the force
density acting in the bulk fluid and we term them ``motility force and
torque'', as they are (minus) the force and torque, respectively, that
should be applied externally on the system ``particle + fluid'' to
keep the particle at rest against its tendency to move. Equivalently,
$\bF_\mathrm{mot}$ and $\bT_\mathrm{mot}$ are responsible for particle
motion when the system is mechanically isolated, i.e., under vanishing
external force and torque on the system:
$\bF^\mathrm{(p)}_\mathrm{ext} + \bF^\mathrm{(f)}_\mathrm{ext} = 0$,
$\bT^\mathrm{(p)}_\mathrm{ext} + \bT^\mathrm{(f)}_\mathrm{ext} = 0$
(an example is the case of electrophoresis, in which an external
electric field acts both on the charged particle and on the ionic
double--layer in the fluid, while the ensemble ``particle +
double--layer'' remains force- and torque-free). When
$\mathbf{f}_\mathrm{ext} \equiv 0$ but
$\mathbf{f}_\mathrm{part} \neq 0$, $\bF_{\mathrm{mot}}$ and
$\bT_{\mathrm{mot}}$ recover the expressions of the ``swimming'' force
and torque introduced by Ref.~\cite{TYB14} (when at least one of them
is non-zero, the particle becomes a ``swimmer'', i.e., it exhibits
``self-motility'' solely through its interaction with the fluid).  The
other limit case, $\mathbf{f}_\mathrm{ext} \neq 0$ but
$\mathbf{f}_\mathrm{part} \equiv 0$, also presents interesting
aspects, in that it corresponds to particle drift by the ambient flow
induced by a force distribution $\mathbf{f}_\mathrm{ext}$ with no net
force and torque on the fluid center of mass
($\bF^{\mathrm{(f)}}_\mathrm{ext} = 0$,
$\bT^{\mathrm{(f)}}_\mathrm{ext} = 0$).

As noted, the physical results should be sensitive only to the
solenoidal component of the force field $\mathbf{f}(\br)$, i.e.,
\eqs{eq:V} and \eref{eq:O} should be invariant with respect to the
transformation $\mathbf{f} \mapsto \mathbf{f} + \nabla \chi$ for any
well behaved and sufficiently fast decaying scalar potential
$\chi(\br)$. The motility force and torque are invariant because any
gradient added to $\mathbf{f}$ drops from \eqs{eq:FTswim} upon
integration by parts and use of the incompressibility
constraints~(\ref{eq:Gincomp}) and of the boundary
conditions~(\ref{eq:GatPart}). (We note in passing that this argument
also rules out self-propulsion based solely on the osmotic pressure of
a solute, in agreement with Refs.~\cite{JuPr09a,JuPr09b,FiDh09}.) On
the other hand, the irrelevance of the potential (longitudinal)
component of $\mathbf{f}$ can be made explicit by expressing the
motility force and torque in terms solely of $\nabla\times\mathbf{f}$,
which renders a formulation with the incompressibility constraint
explicitly accounted for. This can be achieved with the use of the hydrodynamic
potentials: starting from Eqs.~(\ref{eq:potentials}, \ref{eq:AQatPart}), one can
write the motility torque as 
\begin{subequations}
  \label{eq:FTswimAQ}
  \begin{eqnarray}
    \label{eq:TQ}
	\bT_\mathrm{mot} 
    & =
    & \intfluid \bK\times \mathbf{f}
      = \intfluid (-\nabla \Qpot) \times \mathbf{f} \nonumber
      = \intfluid \left[
      \Qpot \, \nabla\times\mathbf{f}
      - \nabla \times (\Qpot \mathbf{f})
      \right]
    \nonumber
    \\
    & =
    &  \intfluid \Qpot \, \nabla\times\mathbf{f}
      - \smashoperator{\oint_{\Spart\cup \mathcal{S}_\infty}}
      d\boldsymbol{\mathcal{S}} \times \Qpot \mathbf{f}
      = \intfluid \Qpot \, \nabla\times\mathbf{f}
      - \phi \smashoperator{\oint_{\Spart}}
      d\boldsymbol{\mathcal{S}} \times \mathbf{f}
      =
      \intfluid \left[ \Qpot - \phi \right] \,\nabla\times\mathbf{f} ,
\end{eqnarray}
where the surface integral over $\mathcal{S}_\infty$ drops due to
the assumption that the field $\mathbf{f}$ vanishes fast enough at
infinity. Likewise, one gets for the 
motility force 
the following alternative expression for each $j=1,2,3$:
\begin{eqnarray}
  \label{eq:FA}
  \be_j \cdot 
  \bF_\mathrm{mot}
  & =
  & \intfluid \be_j\cdot\mathsf{K} \cdot\mathbf{f}
    = \intfluid 
    \left( \nabla\times \bA^{(j)} \right) \cdot \mathbf{f}
    = \intfluid \bA^{(j)}\cdot\left(\nabla\times\mathbf{f}\right)
    + \intfluid \nabla\cdot\left( 
    \bA^{(j)} \times \mathbf{f}
    \right)
    \nonumber
  \\
  & =
  & \intfluid \bA^{(j)}\cdot\left(\nabla\times\mathbf{f}\right)
    + \smashoperator{\oint_{\Spart\cup \mathcal{S}_\infty}}
    d\boldsymbol{\mathcal{S}} \cdot \left(\bA^{(j)} \times \mathbf{f}\right)
    = \intfluid \bA^{(j)}\cdot\left(\nabla\times\mathbf{f}\right)
    + \smashoperator{\oint_{\Spart}}
    d\boldsymbol{\mathcal{S}} \cdot \left(\bA^{(j)} \times
    \mathbf{f} \right)
    \nonumber
  \\
  & =
  & \intfluid \bA^{(j)}\cdot\left(\nabla\times\mathbf{f}\right)
    + \smashoperator{\oint_{\Spart}}
    d\boldsymbol{\mathcal{S}} \cdot \left[
    \nabla \times
    \left( \alpha^{(j)} \mathbf{f} \right)
    - \alpha^{(j)} \, \nabla \times
    \mathbf{f}
    \right]
    \nonumber
  \\
  & =
  & \intfluid \bA^{(j)}\cdot\left(\nabla\times\mathbf{f}\right)
    - \smashoperator{\oint_{\Spart}}
    d\boldsymbol{\mathcal{S}} \cdot \left(\nabla \times
    \mathbf{f} \right) \, \alpha^{(j)} .
\end{eqnarray}
\end{subequations}
In this derivation we have dropped the integral over
$\mathcal{S}_\infty$ again, used the boundary
condition~(\ref{eq:AQatPart}), and, in the last step, taken into
account that the surface $\mathcal{S}_\mathrm{p}$ is closed (it has no
boundary). Now any gradient added to $\mathbf{f}$ is obviously
irrelevant in these expressions for $\bF_\mathrm{mot}$ and
$\bT_\mathrm{mot}$; that arbitrariness has been replaced by the gauge
freedom in the hydrodynamic potentials and, consistently with the
discussion around \eq{eq:AQatPart}, the above expressions are overtly
invariant under a gauge transformation~(\ref{eq:gauge}). One can
alternatively state that \eqs{eq:FTswimAQ} must be evaluated in the
Coulomb gauge, i.e., with those potentials $\hat{\Qpot}$,
$\hat{\bA}^{(j)}$ that verify Eqs.~(\ref{eq:AQatPart},
\ref{eq:Coulomb}). With this choice for the gauge, one arrives at the
simple expressions
\begin{equation}
  \label{eq:FTswimAQHL}
  \bF_\mathrm{mot}
  := \sum_{j=1}^3 \be_j \intfluid
  \hat{\bA}^{(j)}(\br) \cdot \left[ \nabla\times\mathbf{f}(\br)
    \right],
  \qquad
  \bT_\mathrm{mot}   
  := \intfluid \hat{\Qpot}(\br) \,
  \nabla\times\mathbf{f}(\br) .
\end{equation}

While the two expressions for the motility force and torque, namely, Eqs.~(\ref{eq:FTswim}) and (\ref{eq:FTswimAQHL}), are equivalent as long as the exact expressions are employed, when approximations are
required, e.g., for analytical tractability or for numerical
solutions, the use of the explicitly incompressible formulation can be more advantageous (see, e.g., \rcite{DoPo24b}). As an illustration of this point, we consider 
the case of self-chemophoresis, where a particle swims due to the
interaction with self-generated gradients in the chemical composition
of the ambient fluid while in the absence of forces of external origin
(i.e., $\bF^\textrm{(p)}_\textrm{ext} = 0$,
$\bT^\textrm{(p)}_\textrm{ext} = 0$,
$\mathbf{f}_\textrm{ext}(\br) \equiv 0$), see, e.g.,
Refs.~\cite{DSZK47,Ande89,SaSe12,Brown17,DPRD20,DAPA20,ANV22,ShOl24,DoPo24,DoPo24b}.
 The volume force is modeled as \cite{DPRD20} 
\begin{equation}
  \label{eq:ftherm}
  \mathbf{f} = \mathbf{f}_\mathrm{part} = - n \nabla \mu ,
\end{equation}
where $n$ is the concentration of a chemical in the fluid, and $\mu$
is the corresponding chemical potential, which already incorporates
the interaction with the particle. A usual configuration in
experiments occurs when this interaction does not extend much far
apart from the particle, so that its effect is spatially limited to a
layer, which is very thin when compared with the geometrical length
scales associated to the shape of $\Spart$, at the surface of the
particle. This feature can be implemented by approximating the
hydrodynamic kernels in Eqs.~(\ref{eq:FTswim}, \ref{eq:FTswimAQHL}) by
their behavior near this surface, given that they will only depend on
those large geometrical scales. Before considering a particle of
arbitrary shape, for reasons of physically insightful simplicity we
address first the case of a spherical particle of radius $R$. In this
case, analytical expressions for the hydrodynamic kernels $\mathsf{K}$
and $\bK$ are available \cite{HaBr73,KiKa91}:
\begin{equation}
  \label{eq:Kspher}
  \mathsf{K} (\br) =
  \left[ 
    \frac{1}{4} \left(\frac{R}{r}\right)^3
    + \frac{3R}{4r} - 1
  \right] \mathsf{I}
  + \frac{3R}{4r} \left[ 1 - \left(\frac{R}{r}\right)^2 \right]
  \be_r \be_r,
  \qquad
  \bK (\br) = \left[ \left(\frac{R}{r}\right)^3 - 1 \right] r \be_r ,
\end{equation}
in spherical coordinates with origin at the sphere center, where
$\be_r$ is the unit radial vector. The potentials that satisfy the
Coulomb gauge~(\ref{eq:Coulomb}) are given as
\begin{equation}
  \label{eq:AQspher}
  \hat{\bA}^{(j)} (\br) = \frac{3}{4} R \left[ 1 - \frac{2}{3} \frac{r}{R} -
    \frac{1}{3} \left(\frac{R}{r}\right)^2 \right] \be_j\times\be_r ,
  \qquad
  \hat{\Qpot}(\br) = - \frac{3}{2} R^2 \left[ 1 - \frac{2}{3} \frac{R}{r} -
    \frac{1}{3} \left(\frac{r}{R}\right)^2 \right] .
\end{equation}
In the thin--layer approximation, these functions are approximated by
their Taylor expansions in the radial distance $r$ around the
particle's surface ($r=R$), accounting that the fast decay of the force
field $\mathbf{f}(r\be_r)$ at infinity will serve as an effective
cutoff ($r-R\ll R$) in the volume integrals appearing in Eqs.~(\ref{eq:FTswim},
\ref{eq:FTswimAQHL}):
\begin{equation}
  \label{eq:GthinSphere}
  \mathsf{K}(\br) \approx - \frac{3 (r-R)}{2R} \left( \mathsf{I} -
    \be_r \be_r \right) , 
  \qquad
  \bK (\br) \approx - 3 (r-R) \, \be_r,
\end{equation}
and
\begin{equation}
  \hat{\bA}^{(j)}(\br) \approx - \frac{3}{4R} (r-R)^2 \, \be_j\times\be_r,
  \qquad
  \hat{\Qpot}(\br) \approx \frac{3}{2} (r-R)^2 .
\end{equation}
Accordingly, one defines auxiliary fields on the particle's surface as
\begin{equation}
  \label{eq:surfacef}
 \boldsymbol{\mathfrak{f}} (\brp) :=
  \smashoperator{\int_0^\infty} dz\; z\,
  \mathbf{f}_\parallel (\brp+z \bn(\brp)) ,
  \qquad 
  \boldsymbol{\mathfrak{g}} (\brp) :=
  \smashoperator{\int_0^\infty} dz\; z^2\,
  \nabla\times\mathbf{f} (\brp+z \bn(\brp)) ,
\end{equation}
using the notation introduced in \fig{fig:schematic}(right) with
$z:=r-R$, $\bn=\be_r$, $\brp=R \be_r$, so that \eqs{eq:FTswim} become
(here, $d\Omega$ is the element of spherical solid angle and
$\mathbb{S}_2$ is the unit sphere)
\begin{equation}
  \label{eq:FTswimThinSphere}
  \bF_\mathrm{mot}
  \approx - \frac{3R}{2}
  \smashoperator{\int_{\mathbb{S}_2}} d\Omega \;
  \boldsymbol{\mathfrak{f}} (\be_r) ,
  \qquad
  \bT_\mathrm{mot}
  \approx
  - 3 R^2 \smashoperator{\int_{\mathbb{S}_2}} d\Omega \;
  \be_r\times \boldsymbol{\mathfrak{f}} (\be_r) ,
\end{equation}
while \eqs{eq:FTswimAQHL} take the form
\begin{equation}
  \label{eq:FTswimAQHLThinSphere}
  \bF_\mathrm{mot}
  \approx - \frac{3R}{4}
  \smashoperator{\int_{\mathbb{S}_2}} d\Omega \;
  \be_r\times \boldsymbol{\mathfrak{g}} (\be_r) ,
  \qquad 
  \bT_\mathrm{mot}
  \approx \frac{3 R^2}{2}
  \smashoperator{\int_{\mathbb{S}_2}} d\Omega \;
  \boldsymbol{\mathfrak{g}} (\be_r) .
\end{equation}
 (The vectorial identity $\sum_{j=1}^3 \be_j \left(\be_j\times\be_r\right)\cdot\bs
= \be_r\times\bs$, which holds for any vector $\bs$, has been used in the derivation of 
$\bF_\mathrm{mot}$ in this last equation.) The significant difference between these two results is that, unlike \eqs{eq:FTswimAQHLThinSphere}, the expressions appearing in
(\ref{eq:FTswimThinSphere}) are no longer invariant under the replacement $\mathbf{f} \to \mathbf{f} + \nabla\chi$, with
$\chi(\br)$ an arbitrary smooth function that vanishes sufficiently
fast at infinity; for instance, one would get
$\bF_\mathrm{mot} \to \bF_\mathrm{mot} + \delta\bF_\mathrm{mot}$ with
a spurious contribution
\begin{equation}
  \delta \bF_\mathrm{mot}
  :=
  - \frac{3R}{2}
  \smashoperator{\int_{\mathbb{S}_2}} d\Omega \; \nabla_\parallel
  \underbrace{\left[ \smashoperator{\int_0^\infty} dz\; z\, \chi ((R+z)\be_r) \right]}_{\hat{\chi} (\be_r)}
\neq 0 \textrm{ if $\hat{\chi}$ has a non-vanishing dipolar component.}
\end{equation}

The same conclusion holds in the case of a generic particle shape. The
near--particle behavior of the hydrodynamic kernels can be derived
easily \cite{SM}: 
\begin{equation}
  \label{eq:Knear}
  \mathsf{K} (\br) = z \partial_z
  \mathsf{K}_\parallel(\brp) + o(z^2) ,
  \qquad
  \bK(\br) = z \bn(\brp) \partial_z K_z
  (\brp) + o(z^2) ,
\end{equation}
and
\begin{equation}
  \label{eq:AQnear}
  \hat{\bA}^{(j)} (\br) = \frac{1}{2} z^2 \partial_z^2 \hat{\bA}^{(j)} (\brp)
  + o(z^3) ,
  \qquad
  \hat{\Phi}(\br) = \frac{1}{2} z^2 \partial_z^2 \hat{\Phi}(\brp)
  + o(z^3) .
\end{equation}
Correspondingly, in terms of the auxiliary fields~(\ref{eq:surfacef}),
the Eqs.~(\ref{eq:FTswim}) and (\ref{eq:FTswimAQHL}) are approximated
as the following surface integrals, respectively:
\begin{equation}
  \label{eq:FTswimThin}
  \bF_\mathrm{mot}
  \approx
  \smashoperator{\oint_{\Spart}} d\mathcal{S}\;
  \mathsf{K}_\parallel(\brp) \cdot
  \boldsymbol{\mathfrak{f}} (\brp) ,
  \qquad
  \bT_\mathrm{mot}
  \approx
  \smashoperator{\oint_{\Spart}} d\mathcal{S}\;
  \partial_z K_z 
  (\brp) \, \bn(\brp) \times
  \boldsymbol{\mathfrak{f}} (\brp) ,
\end{equation}
and\footnote{According to footnote~\ref{foot:A}, one could also write,
  in a more compact fashion,
  $\bF_\mathrm{mot} = (1/2) \oint d\mathcal{S}\; (\partial_z^2
  \mathsf{A}) \cdot \boldsymbol{\mathfrak{g}}$.}
\begin{equation}
  \label{eq:FTswimAQHLThin}
\bF_\mathrm{mot}
  \approx \frac{1}{2}
  \sum_{j=1}^3 \;\;
  \smashoperator{\oint_{\Spart}} d\mathcal{S}\;
  \partial_z^2 \hat{\bA}^{(j)} (\brp) \, \be_j \cdot
  \boldsymbol{\mathfrak{g}} (\brp) ,
  \qquad
\bT_\mathrm{mot}
  \approx \frac{1}{2} \;\;
  \smashoperator{\oint_{\Spart}} d\mathcal{S}\;
  \partial_z^2 \hat{\Phi} (\brp) \,
  \boldsymbol{\mathfrak{g}} (\brp) .
\end{equation}
The expressions~(\ref{eq:FTswimThin}) are not invariant under the
change $\mathbf{f} \to \mathbf{f} + \nabla\chi$ because the approximate hydrodynamic kernels~(\ref{eq:Knear}) describe a
shear flow along the particle's surface which is compressible on
$\Spart$ (see \eq{eq:uprime}):
\begin{equation}
  \label{eq:localshear}
  \bu(\br=\brp+z \bn)
  = \bV+\bOmega\times\br + z \bgamma(\brp)
  + o(z^2) ,
\end{equation}
with
\begin{equation}
  \bgamma(\brp) :=
  \bV\cdot\partial_z\mathsf{K}_\parallel(\brp) 
  + \bOmega\times\bn (\brp)\, \partial_z K_z (\brp) ,
  \qquad
  \nabla_\parallel\cdot\bgamma \neq 0
  \textrm{ in general}.
\end{equation}
This non-invariance with respect to potential contributions in the
force field $\mathbf{f}$ is a significant inconvenience in that it
prevents carrying out useful approximations, such as replacing the
force $\mathbf{f}$ in \eq{eq:ftherm} by expressions like $\mu\nabla n$
(advantageous for numerical simulations or comparison with
experiments, where the density $n(\br)$ is the field more easily
accessible) or $(\mu\nabla n-n\nabla\mu)/2$, which differ from the
physical force~(\ref{eq:ftherm}) by a dynamically irrelevant additive
gradient. Moreover, it leaves open the possibility of accidentally
carrying osmotic pressure terms into force contributions to motility,
which are obviously spurious (as also pointed out previously
\cite{JuPr09b,FiDh09}). Therefore, \eqs{eq:FTswimThin} are prone to
ambiguities in dealing with thin-film approximations, and it is
preferable to avoid them in favor of \eqs{eq:FTswimAQHLThin}.

In conclusion, the rigid body representation
in terms of the curl of the force field, see Eqs.~(\ref{eq:VOlin},
\ref{eq:FTswimAQHL}), provides a description which is complementary to
the previously derived one in terms of the force field
\cite{Teub82}. It has the advantage of explicitly accounting for the
incompressibility of the flow. For particles of sufficiently simple
shapes, one can get approximate analytical expressions for the kernels
$\mathsf{K}, \bK$, and their potentials $\bA^{(j)},\Qpot$,
respectively: then Eqs.~(\ref{eq:VOlin},
\ref{eq:FTswimAQHL}) provide closed form integral representations, which are well suited
for straightforward, insightful analytical approaches, without the
concern that approximations might violate incompressibility and
  lead to ambiguities in the final expressions. For particles of
generic shapes lacking any special symmetries, the potentials or the
kernels would have to be computed numerically, as it is usually the
case in hydrodynamics. But then, the representation given by
\eqs{eq:FTswimAQHL} has the advantage of a calculation without
constraints, while the calculation of the kernels $\bK$ and
$\mathsf{K}$ is subject to the stringent constraints shown in
\eqs{eq:Gincomp}, without which spurious contributions from the
potential (longitudinal) components of $\mathbf{f}$ would appear in
the rigid body motion. This is particularly challenging in the case
when the force field $\mathbf{f}$ is significant only in a thin layer
near the particle (like, e.g., in many instances of chemophoresis), 
where the magnitude of the kernels $\bK$ and $\mathsf{K}$ is
intrinsically small because they vanish at the surface of the particle
but vary over length scales much larger than the layer thickness. In
this case, discriminating numerically any small spurious components in
$\bK$ and $\mathsf{K}$ becomes technically challenging. On the other
hand, the representation in terms of the potentials $\Qpot$ and
$\bA^{(j)}$ is immune to such issues and solely requires accurate
numerical computations of the curl of $\mathbf{f}$, which is a rather
standard task.

\begin{acknowledgments}
\label{Acknowledgments}
A.D.~and M.N.P.~acknowledge financial support through grants
ProyExcel\_00505 funded by Junta de Andaluc{\'i}a, and
PID2021-126348NB-I00 funded by MCIN/AEI/10.13039/501100011033 and
``ERDF A way of making Europe''. M.N.P.~also acknowledges support from
Ministerio de Universidades through a Mar{\'i}a Zambrano grant.
\end{acknowledgments}

\bibliography{refs_hydro_note.bib}
\newpage 

\appendix

\section{Hydrodynamic fields near the particle}
\label{sec:shearflow}

It is always possible to introduce a local coordinate system near the
particle's surface by translating each point of the surface along its
normal, see \fig{fig:schematic}(right). Any point $\br$ can be
expressed as
\begin{equation}
  \label{eq:coord}
  \br = \brp + z \bn(\brp) ,
  \qquad
   z \geq 0 , \; \brp\in\Spart ,
\end{equation}
where $\bn(\brp)$ is the unit normal to the particle's surface
pointing into the fluid, and the values of the coordinate $z$ are
never taken too large (i.e., compared to the characteristic curvature
radius of the surface), so that this representation remains well
defined. Any vector field $\bQ(\br)$ can be decomposed into normal and
tangential components,
\begin{equation}
  \bQ (\br) = \bn (\brp) Q_z (\br) + \bQ_\parallel (\br) ,
  \quad
  \bQ_\parallel := \mathsf{P}_\parallel\cdot\bQ ,
\end{equation}
in terms of the tensor performing the tangential projection,
\begin{equation}
  \label{eq:P}
  \mathsf{P}_\parallel (\brp) := \mathsf{I} - \bn(\brp)\, \bn(\brp) .
\end{equation}
One can split the differential operator likewise:
\begin{equation}
  \label{eq:nabla}
  \nabla = \bn(\brp) \partial_z + \nabla_\parallel,
  \quad
  \nabla_\parallel := \mathsf{P}_\parallel (\brp)\cdot\nabla .
\end{equation}
The unit normal verifies
\begin{subequations}
  \label{eq:nablan}
\begin{eqnarray}
  \partial_z\bn = \partial_z \mathsf{P}_\parallel
  & =
  & 0,
  \\
  \nabla|\bn|^2 = \left(\nabla_\parallel \bn\right)\cdot\bn
  & =
  & 0 ,
  \\
    \nabla\times\bn = \nabla_\parallel\times\bn
  & =
  & 0 ,
\end{eqnarray}
\end{subequations}
which represent, respectively, that $\bn$ is transported parallel to
itself, that its modulus is unchanged, and that the normal does
  not twist or bend.\footnote{This last equality is derived by applying
    Stokes theorem to the field $\nabla\times\bn$ and noting that
    $d\boldsymbol{\ell}\cdot\bn = dz$ along any path element
    $d\boldsymbol{\ell} = d\br(\brp,z) = d\brp + z\, (d\brp\cdot\nabla_\parallel)
    \bn + \bn \, dz$.}

We first extract the behavior of the fields $\bK$ and $\mathsf{K}$
near the surface, i.e., as an expansion in integer powers of the
coordinate $z$ (here we use the simplified notation $\bk := \be_j\cdot\mathsf{K}$ for each
$j=1,2,3$). One writes
\begin{subequations}
\label{eq:Gthin}
\begin{eqnarray}
  \bk(\br)
  & =
  & \bk(\brp) + z \, \partial_z \bk(\brp)
    + o(z^2) ,
  \\
  \bK(\br)
  & =
  & \bK(\brp) + z \, \partial_z \bK(\brp)
    + o(z^2) .
\end{eqnarray}
\end{subequations}
Due to the no-slip boundary condition~(\ref{eq:GatPart}), one
concludes that the first term in these expansions vanishes. All the
tangential derivatives at the surface also vanish (as derivatives of a
constant),
\begin{equation}
  \nabla_\parallel \bk(\brp) = 0 ,
  \qquad
  \nabla_\parallel \bK(\brp) = 0.
\end{equation}
The latter are employed to evaluate the incompressibility
constraints~(\ref{eq:Gincomp}) at the surface by application of the
identities~(\ref{eq:nablan}):
\begin{equation}
  0 = \nabla\cdot\bk
  = \nabla_\parallel\cdot\bk + \partial_z \left( \bn \cdot \bk \right)
  \quad\stackrel{z=0}{\Rightarrow}\quad
  \partial_z k_z (\brp) = 0 ,
\end{equation}
\begin{equation}
  0 = \nabla\times\bK
  = \nabla_\parallel\times\bK + \partial_z \left(
    \bn \times \bK \right)
  \quad\stackrel{z=0}{\Rightarrow}\quad
  \partial_z \bK_\parallel (\brp) = 0 .
\end{equation}
Therefore, the expansions~(\ref{eq:Gthin}) simplify to
\begin{subequations}
  \label{eq:Gthin2}
  \begin{eqnarray}
    \label{eq:gtransthin}
  \bk(\brp+z \bn)
  & = 
  & z \, \partial_z \bk_\parallel (\brp)
    + o(z^2) ,
  \\
  \label{eq:Grotthin}
  \bK(\brp+z \bn)
  & =
  & z \, \partial_z K_z (\brp) \, \bn (\brp)
    + o(z^2) .
\end{eqnarray}
\end{subequations}

\section{The scalar potential}

Consider the scalar potential defined by \eq{eq:potQ}, for which
\eq{eq:Grotthin} implies the constraint
\begin{equation}
  \label{eq:gradQpart}
  \bK (\br) =-\nabla \Qpot(\br) =0 ,
  \qquad
  \br\in\Spart.
\end{equation}
This has two consequences: the normal derivative vanishes at the surface,
$\partial_z\Qpot(\brp) = K_z= o(z)$, and $\Spart$ is an equipotential,
\begin{equation}
  \label{eq:Qpart}
  \Qpot(\brp) = \phi .
\end{equation}
This boundary condition exhausts the gauge freedom in $\Qpot$, and
allows one to determine it uniquely by integrating \eq{eq:potQ} along
any path connecting the point $\br$ with the particle's surface,
\begin{equation}
  \Qpot(\br) = \phi + \int_{\brp}^\br
  d\boldsymbol{\ell}'\cdot\bK(\br') .
\end{equation}

There is an alternative representation of the scalar potential,
paralleling the representation of the vector potential introduced in
App.~\ref{app:potA}. The velocity field~(\ref{eq:uprime}) can be expressed asymptotically
as a multipolar expansion \cite{KiKa91},
\begin{subequations}
  \label{eq:multipoleExp}
\begin{equation}
  \bu'(\br) = \bu_\infty(\br) + o\left(\frac{1}{r^3}\right) ,
\end{equation}
where the field $\bu_\infty(\br)$ accounts just for the leading terms
in the asymptotic behavior:
\begin{equation}
  \bu_\infty(\br) := \bF \cdot \mathsf{G}(\br)
  + \frac{1}{2} \bT\cdot \left[ \nabla\times\mathsf{G}(\br)
  \right]
  + (\mathsf{S}\cdot\nabla)\cdot\mathsf{G}(\br) ,
\end{equation}
in terms of the Oseen tensor,
\begin{equation}
  \label{eq:oseen}
  \mathsf{G}(\br) := \frac{1}{8\pi \eta r} \left[ \mathsf{I} + \frac{\br
      \br}{r^2} \right],
\end{equation}
\end{subequations}
and the Stokeslet $\bF$, the rotlet $\bT$, and the 
stresslet $\mathsf{S}$, which depend linearly on $\bV', \bOmega'$
through \eqs{eq:rigbod_mobility}. This translates into a specific
asymptotic behavior of $\bK(\br)$: by considering a purely rotational
motion ($\bV'=0$ in \eq{eq:uprime}), 
one can write
\begin{equation}
  \bK (\br) = \bK_{\infty}(\br) + \delta\bK(\br) ,
\end{equation}
where $\bK_{\infty}(\br)$ is defined through the equality
\begin{equation}
  \label{eq:Kinfty}
  \bu_\infty(\br) =: \bOmega\times \left[ \br + \bK_\infty(\br) \right] ,
\end{equation}
so that
\begin{equation}
  \label{eq:deltaK}
  \delta\bK(\br) = o\left(\frac{1}{r^3}\right)
  \;\mathrm{as}\; |\br|\to \infty .
\end{equation}
One can decompose the scalar potential likewise as
\begin{equation}
  \Qpot(\br) = \Qpot_{\infty}(\br) + \delta\Qpot(\br) ,
\end{equation}
with
\begin{equation}
  \bK_\infty = - \nabla \Qpot_\infty,
  \qquad
  \delta\bK = - \nabla \delta\Qpot .
\end{equation}
At this stage, $\bK_{\infty}(\br)$ and $\Qpot_\infty$ are known while,
by \eq{eq:deltaK}, the scalar potential $\delta\Qpot$ vanishes
sufficiently fast at infinity that one can apply Green's identity in
order to represent it as (recalling that $\Spart$ is oriented inwards
to $\mathcal{D}$)
\begin{equation}
  \delta\Qpot(\br) = \frac{1}{4\pi} \smashoperator{\int_{\mathcal{D}}} d\mathcal{V}'\;
  \frac{\nabla'\cdot\delta\bK(\br')}{|\br-\br'|}
  + \frac{1}{4\pi} \smashoperator{\oint_\Spart}
  d\boldsymbol{\mathcal{S}'}
  \cdot\left[
    \delta\Qpot(\br') \nabla'\left(\frac{1}{|\br-\br'|} \right)
    - \frac{\nabla' \delta\Qpot(\br')}{|\br-\br'|}
  \right] ,
\end{equation}
in terms of the boundary conditions~(\ref{eq:gradQpart},
\ref{eq:Qpart}):
\begin{equation}
  \delta\Qpot(\br) = \phi - \Qpot_\infty(\br) ,
  \quad
  \nabla\delta\Qpot(\br) = - \nabla\Qpot_\infty(\br) ,
  \qquad
  \br\in\Spart.
\end{equation}

Yet another alternative to determine the scalar potential 
exploits that it is the (unique) solution of the following
(electrostatic) boundary--value problem:
\begin{subequations}
\begin{equation}
  \nabla^2\Qpot(\br) = - \nabla\cdot\bK (\br) ,
  \qquad
  \br\in\mathcal{D} ,
\end{equation}
\begin{equation}
  \Qpot(\br) = \phi ,
  \qquad
  \br\in\Spart ,
\end{equation}
\begin{equation}
  \Qpot(\br) \sim \Qpot_{\infty}(\br) ,
  \qquad
  \mathrm{as}\; |\br|\to\infty .
\end{equation}
\end{subequations}
The solution of this problem for the value $\phi=0$
in the case of a spherical particle leads to the expression for
$\hat{\Qpot}$ shown in~(\ref{eq:AQspher}).

Finally, we note that the conditions~(\ref{eq:gradQpart},
\ref{eq:Qpart}) lead straightforwardly to the following near-particle
behavior for the scalar potential in the Coulomb gauge given by
\eq{eq:Coulomb}:
\begin{equation}
  \label{eq:Qthin}
  \hat{\Qpot}(\brp+z \bn) = \frac{1}{2} z^2 \partial_z^2 \hat{\Qpot}(\brp)
  + o(z^3) .
\end{equation}

\section{The vector potentials}
\label{app:potA}

The vector potentials are defined by \eq{eq:potA}. As in
App.~\ref{sec:shearflow}, we use the simplified notation
$\bA := \bA^{(j)}$ for each $j=1,2,3$, with the definition
\begin{equation}
  \nabla\times\bA (\br) = \bk(\br) \quad (:= \be_j\cdot\mathsf{K}(\br) ).
\end{equation}
In view of \eq{eq:gtransthin}, the field $\bA$ is conservative up to
$o(z)$ near the surface of the particle. More specifically, with the
general ansatz
\begin{equation}
  \bA(\brp + z \bn) = \nabla \alpha(\brp + z \bn) + z
  \boldsymbol{\beta}(\brp) + o (z^2)
  \quad\Rightarrow\quad
  \nabla\times\bA(\brp + z \bn)
  = \bn(\brp)\times\boldsymbol{\beta}(\brp) + o(z) ,
\end{equation}
one concludes, by comparison with \eq{eq:gtransthin}, that
$\boldsymbol{\beta}(\brp)$ is an arbitrary normal vector field, i.e.,
\begin{equation}
  \label{eq:AnearPart}
  \bA(\brp + z \bn) = \nabla \alpha(\brp + z \bn) + z \bn
  \beta(\brp) + o(z^2) .
\end{equation}
One can recognize in $\alpha(\brp+z\bn)$ an arbitrary scalar field
that changes the gauge, i.e., the value
$\Theta (\br\in\mathcal{D}) := \nabla\cdot\bA(\br)$ of the divergence
of the vector potential. Turning the argument around, one can impose
that the field $\alpha$ is determined by the bulk field $\Theta(\br)$,
by the surface field $\alpha_p(\brp) :=
\alpha(\brp)$, 
and by vanishing at infinity, that is, the field $\alpha$ is the
unique solution of the following boundary--value problem:
\begin{subequations}
  \label{eq:alpha}
  \begin{equation}
    \nabla^2\alpha(\br) = \Theta(\br) ,
    \qquad
    \br\in\mathcal{D} ,
  \end{equation}
  \begin{equation}
    \label{eq:alphap}
    \alpha(\br) = \alpha_p(\br),  
    \qquad
    \br\in\Spart,
  \end{equation}
  \begin{equation}
    \label{eq:alphainfty}
    \alpha(\br) \to 0 ,
    \qquad
    \mathrm{as}\; |\br|\to\infty .
  \end{equation}
\end{subequations}
(This latter behavior at infinity, together with a similar constraint
on the behavior of $\Theta(\br)$, excludes gauges fixed by external 
sources foreign to specific features of the particle itself, and yield a
well--posed problem.) An immediate consequence is that the field
$\beta(\brp)$ must vanish: by using the
decomposition~(\ref{eq:AnearPart}), one gets
\begin{equation}
  \label{eq:beta}
  \Theta (\br) := \nabla\cdot\bA(\br) = \nabla^2 \alpha(\brp + z \bn) + \beta(\brp) + o(z)
  \quad\Rightarrow\quad
  \beta(\brp) = 0 .
\end{equation}
Thus, the vector potential at the particle's surface takes a simple
form
\begin{equation}
  \label{eq:AatPart}
  \bA(\br) = \nabla\alpha(\br) ,
  \qquad
  \br\in\Spart ,
\end{equation}
and the boundary condition~(\ref{eq:alphap}) is equivalent to
specifying the tangential component $\bA_\parallel(\brp)$ of the
vector potential on the surface of the particle. (But one could have
instead specified the normal component $A_z(\brp)$ on the surface,
which would correspond to a different gauge, namely, the solution of
the above boundary--value problem with a prescribed surface field
$\partial_z\alpha(\brp)$.) Finally, we notice that
$\alpha(\br)\equiv 0$ is a valid gauge, which we call ``Coulomb
gauge'' because $\Theta(\br)\equiv 0$, and in which $\bA$ vanishes
altogether at the particle's surface.

In order to obtain a well-posed problem for $\bA(\br)$, one has to
address also the boundary conditions at infinity, which are determined
by the multipolar expansion~(\ref{eq:multipoleExp}). We thus proceed as with
the scalar potential but with the Stokeslet, rotlet, and stresslet
determined by a purely translational motion ($\bOmega'=0$ in
\eq{eq:uprime}): one thus writes
\begin{equation}
  \bk (\br) =
  \bk_{\infty}(\br) + \delta\bk(\br) ,
\end{equation}
with
\begin{equation}
  \label{eq:deltak}
  \delta\bk(\br) = o\left(\frac{1}{r^3}\right)
  \;\mathrm{as}\; |\br|\to \infty ,  
\end{equation}
and the field
\begin{equation}
  \label{eq:kinfty}
  \bk_{\infty}(\br) :=
  -\be_j - \bF \cdot \mathsf{G}(\br)
  + \frac{1}{2} \bT\cdot \left[ \nabla\times\mathsf{G}(\br)
  \right]
  + (\mathsf{S}\cdot\nabla)\cdot\mathsf{G}(\br) .
\end{equation}
Accordingly, one also decomposes the vector potential as
\begin{equation}
  \bA(\br) = \bA_{\infty}(\br) + \delta\bA(\br) ,
\end{equation}
with the field 
\begin{equation}
  \label{eq:Ainfty}
  \bA_{\infty}(\br) := \frac{1}{2} \br\times\be_j +
  \frac{1}{8\pi\eta r}\br\times\bF
  - \frac{1}{2} \mathsf{G}(\br)\cdot\bT
  + \sum\limits_{n,m = 1}^{3}\mathsf{G}_{nm}(\br)
  \be_n\times (\be_m\cdot\mathsf{S}),
\end{equation}
that obeys
\begin{equation}
  \label{eq:fieldEqsAinfty}
  \nabla\times\bA_{\infty} (\br)= \bk_{\infty}(\br),
  \qquad
  \nabla\cdot\bA_{\infty} (\br) = 0 ,
  \qquad
  \br\in\mathcal{D} .
\end{equation}
Therefore, the field $\delta\bA$ satisfies
\begin{subequations}
\begin{equation}
  \label{eq:fieldEqsDeltaA}
  \nabla\times\delta\bA (\br) = \delta\bk,
  \qquad
  \nabla\cdot\delta\bA (\br) = \Theta (\br)
  \;\;(= \nabla^2\alpha(\br) ),
  \qquad
  \br\in\mathcal{D} ,
\end{equation}
\begin{equation}
  \label{eq:deltaAatPart}
  \delta\bA(\br) = \nabla\alpha(\br) - \bA_\infty(\br) ,
  \qquad
  \qquad
  \br\in\Spart.
\end{equation}
\end{subequations}
We further impose the condition that $\delta\bA(\br)$ vanishes at
infinity sufficiently fast for the same reason as argued regarding
\eq{eq:alphainfty}. This constraint together with the fast decay at infinity of
$\delta\bk(\br)$, see \eq{eq:deltak}, warrants that a Helmholtz
decomposition holds for the field $\delta \bA(\br)$; that is, one can
write
\begin{subequations}
\begin{equation}
  \delta\bA(\br) = - \nabla \Psi(\br) + \nabla\times\bB(\br) ,
  \qquad
  \nabla\cdot\bB(\br) = 0 ,
\end{equation}
with the auxiliary fields (recall that $\Spart$ is oriented inwards to
$\mathcal{D}$)
\begin{equation}
  \Psi(\br) = \frac{1}{4\pi} \smashoperator{\int_{\mathcal{D}}} d\mathcal{V}'\;
  \frac{\Theta(\br')}{|\br-\br'|}
  + \frac{1}{4\pi} \smashoperator{\oint_\Spart}
  d\boldsymbol{\mathcal{S}}'\cdot\frac{\delta\bA(\br')}{|\br-\br'|} ,
\end{equation}
\begin{equation}
  \bB(\br) = \frac{1}{4\pi} \smashoperator{\int_{\mathcal{D}}} d\mathcal{V}'\;
  \frac{\delta \bk(\br')}{|\br-\br'|}
  + \frac{1}{4\pi} \smashoperator{\oint_\Spart}
  d\boldsymbol{\mathcal{S}}'\times\frac{\delta\bA(\br')}{|\br-\br'|} ,
\end{equation}
\end{subequations}
the gauge freedom being contained in the scalar field
$\alpha(\br\in\mathcal{D})$ via the bulk field $\Theta(\br)$ and the
surface field $\delta\bA(\br')$. In particular, in the Coulomb gauge
($\alpha(\br)=0$), the vector potential can then be represented as
\begin{subequations}
\begin{equation}
  \hat{\bA}(\br) = \bA_\infty(\br) - \nabla \hat{\Psi}(\br) + \nabla\times\hat{\bB}(\br) ,
\end{equation}
with the fields
\begin{equation}
  \hat{\Psi}(\br) = - \frac{1}{4\pi} \smashoperator{\oint_\Spart}
  d\boldsymbol{\mathcal{S}}'\cdot\frac{\bA_\infty(\br')}{|\br-\br'|} ,
\end{equation}
\begin{equation}
  \hat{\bB}(\br) = \frac{1}{4\pi} \smashoperator{\int_{\mathcal{D}}} d\mathcal{V}'\;
  \frac{\delta \bk(\br')}{|\br-\br'|}
  - \frac{1}{4\pi} \smashoperator{\oint_\Spart}
  d\boldsymbol{\mathcal{S}}'\times\frac{\bA_\infty(\br')}{|\br-\br'|} \,;
\end{equation}
\end{subequations}
alternatively (but still in the Coulomb gauge), the vector potential
can be obtained as the (unique) solution of the following
(magnetostatic) boundary-value problem:
\begin{subequations}
\begin{equation}
  \nabla\times\hat{\bA}(\br) = \bk (\br) ,
  \qquad
  \nabla\cdot\hat{\bA}(\br)= 0 ,
  \qquad
  \br\in\mathcal{D} ,
\end{equation}
\begin{equation}
  \hat{\bA}_\parallel(\br) = 0 ,
  \qquad
  \br\in\Spart ,
\end{equation}
\begin{equation}
  \hat{\bA}(\br) \sim \bA_{\infty}(\br) ,
  \qquad
  \mathrm{as}\; |\br|\to\infty .
\end{equation}
\end{subequations}

In the case of a generic shape, one can extract the near--particle
behavior of the vector potential in the Coulomb gauge easily: since
$\bA=o(z^2)$ by Eqs.~(\ref{eq:AnearPart}, \ref{eq:beta}), then
\begin{equation}
  \label{eq:Athin}
  \hat{\bA}(\brp+z \bn) = \frac{1}{2} z^2 \partial_z^2 \hat{\bA}(\brp)
  + o(z^3) .
\end{equation}

\end{document}